\UseRawInputEncoding
\documentclass[aps, prb, letterpaper, 10pt, twocolumn, superscriptaddress]{revtex4-2}

\usepackage[colorlinks=true,allcolors=blue!80!black]{hyperref}

\usepackage{amsmath}
\usepackage[caption = false]{subfig}
\usepackage{graphicx,epstopdf}
\usepackage{blindtext}
\usepackage{lipsum}
\usepackage{amsfonts}
\usepackage{bbm}
\usepackage{amssymb}
\usepackage{enumerate}
\usepackage{color}
\usepackage{latexsym}
\usepackage{physics}
\usepackage{times,txfonts}
\usepackage{xcolor}
\usepackage{epsfig}
\usepackage{bm}
\usepackage{tabularx}
\usepackage{multirow}
\usepackage{mathtools}
\usepackage{xfrac}
\usepackage{physics}
\usepackage{enumerate}
\usepackage{braket}

\newcommand{\proj}[1]{\ket{#1}\!\bra{#1}}
\newcommand{\trans}[2]{\ket{#1}\!\bra{#2}}
\newcommand{\mean}[1]{\langle #1 \rangle}
\newcommand{\one}{\openone}

\renewcommand{\tr}[2]{\mathrm{Tr}_{#1}\left\{#2\right\}}
\newcommand{\hatd}[1]{\hat{#1}^{\dagger}}

\newcommand{\half}{\frac{1}{2}}

\newcommand{\vnnb}{V$_{\rm B}$}

\newcommand{\gtxt}[1]{{\color{green} #1}}

\begin{document}
\title{Active spin lattice hyperpolarization: Application to hexagonal boron nitride color centers}

\author{F. T. Tabesh}
\email{fatemeh.tabesh@gmail.com}
\affiliation{Department of Physics, Isfahan University of Technology, Isfahan 84156-83111, Iran}
\author{M. Fani}
\affiliation{Department of Physics, Isfahan University of Technology, Isfahan 84156-83111, Iran}
\author{J. S. Pedernales}
\affiliation{Institut f{\"u}r Theoretische Physik, Albert-Einstein-Allee 11, Universit{\"a}t Ulm, 89069 Ulm, Germany}
\author{M. B. Plenio}
\affiliation{Institut f{\"u}r Theoretische Physik, Albert-Einstein-Allee 11, Universit{\"a}t Ulm, 89069 Ulm, Germany}
\author{M. Abdi}
\email{mehabdi@gmail.com}
\affiliation{Department of Physics, Isfahan University of Technology, Isfahan 84156-83111, Iran}

\date{\today}
\begin{abstract}
The active driving of the electron spin of a color center is known as a method for the hyperpolarization of the surrounding nuclear spin bath and to initialize a system with large number of spins.
Here, we investigate the efficiency of this approach for various spin coupling schemes in a one-dimensional Heisenberg chain coupled to a central spin.
To extend our study to the realistic systems with a large number of interacting spins, we employ an approximate method based on Holstein-Primakoff transformation.
The validity of the method for describing spin polarization dynamics is benchmarked by the exact numerics for a small lattice, where the accuracy of the bosonic Holstein-Primakoff approximation approach is confirmed.
We, thus, extend our analysis to larger spin systems where the exact numerics are out of reach.
The results prove the efficiency of the active driving method when the central spin interaction with the spin bath is long range and the inter-spin interactions in the bath spins is large enough.
The method is then applied to the realistic case of optically active negatively charged boron vacancy centers (\vnnb) in hexagonal boron nitride.
Our results suggest that a high degree of  hyperpolarization in the boron and nitrogen nuclear spin lattices  is achievable even starting from a fully thermal bath.
As an initialization, our work provides the first step toward the realization of a two-dimensional quantum simulator based on natural nuclear spins and it can prove useful for extending the coherence time of the \vnnb~centers.

\end{abstract}

\maketitle

%
%
\section{Introduction}%
Various properties of defect centers in solid-state materials have been studied considerably for a long time~\cite{stoneham2001theory}, especially Nitrogen-Vacancy (NV) centers in diamond~\cite{Doherty2013} (and references therein) and different defects in silicon carbide~\cite{koehl2011, widmann2015}. In particular, it has been identified that electron spins of defect centers in wide-band gap semiconductors, most notably diamond, can be initialized optically and controlled by microwaves~\cite{Doherty2013}. In addition, the controlled coupling of these electron spins with proximal nuclear spins (of e.g. nitrogen atoms for the NV center cases or $^{13}$C) have been achieved by using microwave pulses~\cite{dolde2014high}, electrical~\cite{pla2013high}, or optical detection~\cite{jelezko2004observation}.

On the other hand, extended systems such as nuclear spins on the surface of diamond~\cite{Cai2013} or  thin $^{13}$C layers in diamond~\cite{unden2018coherent} have been recently proposed as potential quantum simulators. However, a key challenge of these implementations is the initialization of such nuclear spin ensemble, i.e. generation of a robust hyper-polarized state with nearly 100$\%$ spin polarization~\cite{Cai2013,unden2018coherent}.  Therefore, various methods have been studied  to achieve this high level of spin polarization by employing color centers in diamond~\cite{Cai2013,London2013,Rej2015,Healey2021,Fernandez2018a,alvarez2015local,king2015room,Scheuer2016} and in silicon carbide~\cite{Falk2015}. The highly controllable color centers can be polarized efficiently at room temperature via optical and microwave drives, then their polarization is transferred to
other interacting spin species.

In spite of the various valuable works on NV centers in diamond,
spin defects in non-carbon lattices have mostly been overlooked,
while there are tremendous unexplored areas outside
the carbon realm. More recently, defect centers in 2D materials such as hexagonal boron nitride (hBN) have been identified experimentally~\cite{Tran2016} and characterized theoretically~\cite{Sajid2020,Abdi2018}. In this case, due to the simultaneous presence of different nuclear spin species in a lattice structure, the initialization of spin ensemble is more complex. Primary experimental attempts to initialize the spin ensemble in hBN based on the anti-crossing levels have been recently reported~\cite{gao2022nuclear}. In this paper, we adopt approaches developed in the field of color centers in diamond based on the microwave control of color centers, to examine scalabale schemes for the hyper-polarization of the nuclear spins in hBN.

Here, we study the hyperpolarization of nuclear spins (Borons and Nitrogens) in a mono-layer of hBN lattice via electromagnetic manipulation of the electron spin of \vnnb.
As an immediate application, this can significantly decrease the pure dephasing contribution of the spin bath, and thus, enhancing coherence time of the defect spin state.
A longer coherence time shall prove useful in every follow-up quantum technological applications, see e.g.~\cite{Tabesh2021}.
Unlike the hyperpolarization of the nuclear spins in diamond, here one should deal with the polarization of two sub-lattices with different nuclear species and different spin values.
We survey the hyperpolarization of the hBN lattice by optical pumping and microwave driving and finds its rapid and efficient performance well-beyond the low temperature and high magnetic field levels.
In particular, we examine the direct polarization swap between the \vnnb\ defect and the surrounding nuclei in such a way that a microwave field is applied to handle the electron spin of the \vnnb\ defect. 
In this scheme, the population transfer takes place when the Rabi frequency of the microwave driving field is resonant with the energy splitting of the nuclear spins.
Therefore, the flip-flop processes between the \vnnb\ defect and the nuclear spins can result-in the polarization of the nuclear spin lattice.
Moreover, we investigate the optimal control over the nuclei through adjustment of the magnetic field orientation as well as frequency and amplitude of the microwave drive that excites the electron spin.

In order to corroborate our study with numerical analyses on such large spin systems, we employ an approximate numerical method that overcomes the typical limitations in computational resources for dealing with large scale spin systems.
The numerical method is based on using a bosonic approximation through the Holstein-Primakoff (HP) transformation~\cite{Holstein1940}. 
We justify validity of the approximation in the working regime of our interest by benchmarking its results with the exact numerics for small sized lattices. The study is then extended to the larger lattices with faster and yet considerably less computational resources.
The HP transformation has been used for investigating  hyperpolarization of oil molecules~\cite{Fernandez2018a}, sensing phases of water via NV centers~\cite{Fernandez2018b}, and simulating on a diamond surface~\cite{Cai2013}.

Before addressing the hBN problem, we explore the abilities and limitations of the employed numerical bosonic method and examine the range of its validity by applying it to the simple model of inhomogeneous central spin model.
In this toy model, the spins in a one-dimensional Heisenberg chain inhomogeneously interact with a distinct spin, the `central spin'. 
We argue that the bosonic approach is in good agreement with exact numerical simulation if the interaction between the central spin and the spin bath is long range. Also, we find that the interaction among bath spins plays an effective role to transfer polarization throughout the one-dimensional lattice.
Therefore, we introduce the Heisenberg model and the Gaussian state method in Sec.~\ref{secIsing}. In this example, we are interested in the study of possibility of strong polarization by driving the central spin. In addition, we compare the results of the exact solution and bosonic approximation for a few nuclei. In Sec.~\ref{hBN}, we introduce the Hamiltonian of hBN lattice with a negatively charged \vnnb~ defect and the nuclear bath model. We also study the hyperpolarization in this lattice.  Finally, we summarize and conclude in Sec.~\ref{seccon}.
\begin{figure}[tb]
\includegraphics[width=\columnwidth]{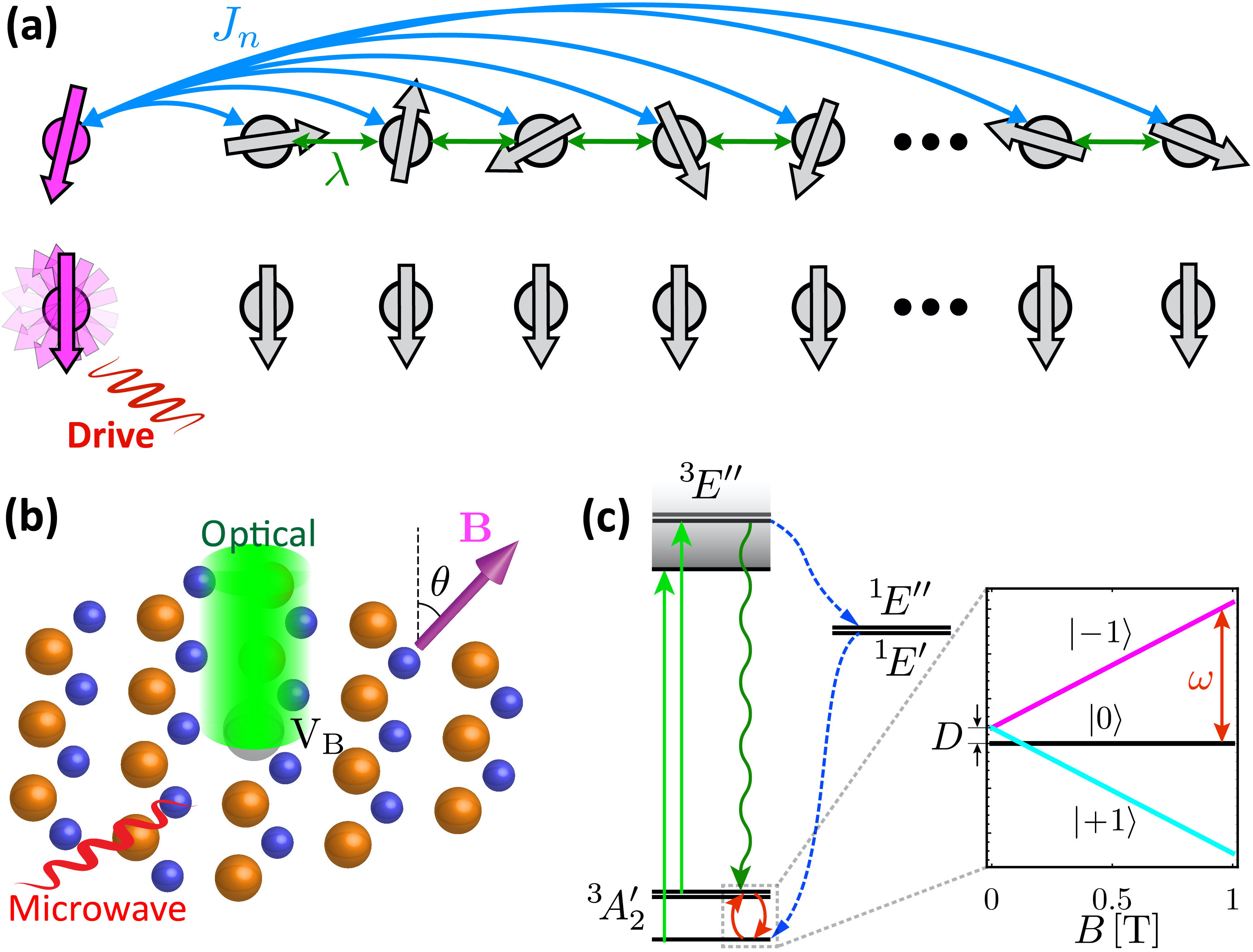}
\caption{%
(a) Sketch of the central spin model studied in this work (top panel) and the hyperpolarization of the bath spin chain as a result of active manipulation of the control spin (bottom panel).
(b) The geometry of \vnnb\ defect in hBN: A negatively charged boron vacancy (gray) surrounded by nitrogen (blue) and boron (orange) atoms.
The optical polarizing and microwave drives are also indicated.
(c) \vnnb\ defect simplified energy level diagram: The straight green lines show the exciting laser transitions, while the curly green and dashed blue lines denote the radiative and non-radiative decay to the ground state, respectively.
The red circle arrows represents the microwave drive.
The inset presents a closer look at the ground state manifold and its manipulation via external magnetic field and microwave drive.
}%
\label{Fig3}%
\end{figure}%
%
%
%
\section{Central spin model}\label{secIsing}
In order to better understand the hyperpolarization process through active manipulation of a defect spin and to examine the area of validity of the employed approximate numerical method, here we investigate the familiar problem of central spin model.

\subsection{Hamiltonian}
We begin with the nearest neighbor Heisenberg model for a chain of one-half spins~\cite{LIEB1961, Baxter, nolting} described by the Hamiltonian $(\hbar=1)$
\begin{equation}
\hat H_{\rm B}= h\sum_{n=1}^N \hat{S}^z_n +\lambda \sum_{n=1}^{N-1} (\hat{S}^z_n\hat{S}^z_{n+1}+\hat{S}^+_n\hat{S}^-_{n+1}+ \hat{S}^-_n\hat{S}^+_{n+1})\text{ , }
\label{i1}
\end{equation}
where $\hat{S}^z_n=\frac{1}{2}\sigma^z_n$ denotes the spin operator in  $z$-direction and $\hat{S}^\pm_n$ is raising/lowering operator on site $n$.
Here, $h$ is the Larmor frequency which is proportional to the background magnetic field and $\lambda$ is the coupling strength.
The first term of Hamiltonian (\ref{i1}) is the free energy of spins and the second term describes the nearest neighbor one-dimensional interactions among the spins.
At zero temperature and in the limit of $\lambda \ll h$, the  ground state of the system is unique and given by $\ket{\mathbf{0}}=\ket{0}^{\otimes N}$, where $\ket{0}$ and $\ket{1}$ are the eigenstates of $\sigma^z$ (Pauli matrix in the  $z$-direction) respectively corresponding to $-1$ and $+1$.

Now, we consider a central spin one-half particle as the control quantum entity, whose free dynamics is described by the Hamiltonian $\hat{H}_{\rm cs}=\omega_0\hat{s}^z$ with Larmor frequency $\omega_0$. 
In a possible realistic physical implementation, the control spin is an optically active spin that a resonant driving field is applied for its manipulation, see  Fig.~\ref{Fig3}(a).
The central (control) spin interacts with the Heisenberg chain through the following Hamiltonian
\begin{equation}
\hat H_{\rm int}=\sum_{n=1}^N J_n\hat{s}^x\hat{S}^x_{n}\text{ , }
\label{i2}
\end{equation}
where $J_n$ are the long-range coupling strengths, while $\hat{s}^z$ and $\hat{s}^x$ are the spin operators of the control spin.
The whole system dynamics is thus given by the Hamiltonian $\hat{H} = \hat{H}_{\rm cs}+\hat H_{\rm B}+\hat{H}_{\rm int}$. 

For the sake of simplicity, we assume that $\lambda,J_n \ll h, \omega_0$.
Hence, the rotating wave approximation (RWA) is safely satisfied and the total Hamiltonian reads
\begin{align}\label{i3}
\hat H &=\omega_0\hat{s}^z +h\sum_{n=1}^N \hat{S}^z_n +\lambda \sum_{n=1}^{N-1}  (\hat{S}^z_n\hat{S}^z_{n+1}+\hat{S}^+_n\hat{S}^-_{n+1}+ \hat{S}^-_n\hat{S}^+_{n+1}) \nonumber\\
&+\sum_{n=1}^N  J_n (\hat{s}^+\hat{S}^-_{n} +\hat{s}^-\hat{S}^+_{n})
\text{ . }
\end{align}
The second line of the above Hamiltonian implies the flip-flop interactions between the control spin and Heisenberg chain spins.


\subsection{Hyperpolarization}\label{sec:isinghyper}
The hyperpolarization is attainable by performing cycles of polarization transfer that is made up of the following steps:
(i) The control spin is initialized in the $\ket{0}$ state in each cycle realization. This can be done, e.g. by applying a laser pulse.
(ii) By bringing the control spin into resonance with the `bath' spins the Hartmann-Hahn condition is satisfied, and thus, the polarization of the control spin is transferred to the bath spins~\cite{Hartmann}.
The frequency tuning of the control spin for satisfying the resonance condition can be done in various ways.
One possibility is to apply a resonant driving field with a proper Rabi frequency, see Sec.~\ref{hBN}.
The two above polarizing steps are repeated as many as necessary times to achieve the desired level of polarization in the Heisenberg chain spins.
To put it in the mathematical form, the total initial state is given by $\rho(0)=\proj{0}_{\rm cs}\bigotimes^{N}_{n=1}\rho^{\rm th}_{n}$ where $\rho_n^{\rm th}$ is the initial unpolarized thermal state of the $n$th spin in the chain.
After the $(R+1)$th iteration of the polarization cycle, the state of the Heisenberg spins is found by
\begin{align}
\hat{\rho}_{\rm B}\big((R+1)\tau\big)=\tr{\rm cs}{\hat{U}(\tau)\big(\proj{0}_{\rm cs}\otimes \hat{\rho}_{\rm B}(R\tau)\big) \hat{U}^\dag(\tau)} \text{, }
\label{i8}
\end{align}
where $\hat{U}(t)\equiv \exp\{-i\hat H t\}$ is the time evolution operator with $\hat H$ the Hamiltonian in Eq.~\eqref{i3} and $\tr{\rm cs}{}$ indicates the partial trace over the control spin.
Here, the period of all cycles is taken identical and equal to $\tau$.
We must emphasize that the polarization can be effectively conveyed from the control spin to the Heisenberg spins when $\omega_0=h$.
Thus, after a perfect hyperpolarization procedure the final state of the bath spins approaches to $\ket{\mathbf{0}}=\ket{0}^{\otimes N}$.
 
In the following we shall perform exact and approximate numerical analysis to study this procedure.
In order to quantify the level of lattice polarization, we define the average collective expectation value of the operators $\hat{S}^{z}$ as
\begin{align}
\overline{S}_{\!z} &=\frac{1}{N}\sum_{n=1}^{N}  \frac{\mean{\hat{S}^{z}_n}}{s_{n}}\text{ , }
\label{i6}
\end{align}
with $\mean{\hat{S}^{z}_n}$ and $s_{n}$ denoting the expectation value and amount of the spin of the $n$th bath spin in the chain, respectively.
In the current case we have $s_{n}=1/2$.
Note that the total polarization is normalized to unity and $-1\leq \overline{S}_{\!z} \leq +1$.
The exact numerics are performed by the QuTiP package~\cite{Johansson2012}.
Nonetheless, due to the limited computational power the hyperpolarization of large spin systems is studied by a method based on the Holstein-Primakoff approximation, which is discussed next.
\subsection{Gaussian states method}
In order to model the behavior of large spin baths, we now employ the Holstein-Primakoff transformation (HPT) and the corresponding approximation (HPA) to compute the time evolution of Eq.~\eqref{i8} for large number of spins based on bosonic states~\cite{Holstein1940, Fernandez2018a}.
In this method, a highly polarized spin is treated as a boson close to its ground state.
It is worth mentioning that despite this fact, in the hyperpolarization problem one usually takes the initial state of the spin bath in a fully thermal state.
Nevertheless, since state of the bath spins get closer to their respective ground state after each hyperpolarization cycle the amount of total error committed in this method is tractable.
On the other hand, bosonic fields whose dynamics is described by Hamiltonians quadratic in the creation and annihilation operators of such fields preserve their Gaussian character in all future times, provided they are initially in a Gaussian state.
Given the fact that Gaussian states are completely characterized by their first and second moments, the equations of motion for those moments is enough for describing the system dynamics.
Notably, those equations of motion have a dimensionality that grows linearly with the number of constituents.
Therefore, their simulation has considerably less complexity compared to the original dynamics.

The Hamiltonian \eqref{i3} can be exactly mapped into bosons under the HPT. However, the resulting bosonic Hamiltonian is not quadratic in the creation and annihilation operators.
This makes it difficult to study the system dynamics. One, therefore, takes its linear approximation, the lowest order of the HPA, that works the best for spin states that are close to their ground state.
In this approximation, the spin operators of a spin-$s$ particle are transformed as the following
\begin{align}\label{HP}
\hat{s}^z &= \hat{a}^\dag\hat{a}-s \one \text{ , }\nonumber\\
\hat{s}^+ &= \hat{a}^\dag \sqrt{2s - \hat{a}^\dag\hat{a}} \approx \hat{a}^\dag \sqrt{2s}\text{ , }\\\nonumber
\hat{s}^- &= \sqrt{2s - \hat{a}^\dag\hat{a}}\ \hat{a} \approx \sqrt{2s}~\hat{a}\text{ . }
\end{align}
where $\hat{a}^\dag$ ($\hat{a}$) is the bosonic creation (annihilation) operator with commutator $[\hat{a},\hatd{a}]=\one$.
Similarly, we perform the transformation for the Heisenberg chain operators by assigning the bosonic operators $\hat{b}_n$ who satisfy the commutation relation $ [\hat{b}_n,\hat{b}^\dag_m]= \one \delta_{nm}$. 
By applying the above transformations the Hamiltonian in Eq.~\eqref{i3} reads
\begin{align}\label{i9}
\hat H(t) &=\omega_0 \hat{a}^\dag\hat{a}+\dfrac{\lambda}{2} (\hat{b}^\dag_1 \hat{b}_1+\hat{b}^\dag _N \hat{b}_N)+(h-\lambda )\sum_{n=1}^N  \hat{b}^\dag_n \hat{b}_n \nonumber\\
&+\lambda\sum_{n=1}^{N-1} \hat{b}^\dag_n \hat{b}_{n+1}+ \hat{b}_n \hat{b}^\dag_{n+1} +\frac{1}{2}\left[ \hat{b}^\dag_n \hat{b}_{n}\overline{n}_{n+1}(t)+\hat{b}^\dag_{n+1} \hat{b}_{n+1}\overline{n}_{n}(t)\right] \nonumber\\
&+\sum_{n=1}^N  J_n(\hat{a}^\dag \hat{b}_{n} +\hat{a} \hat{b}^\dag_{n}) \text{ . }
\end{align}
where to maintain the quadratic form of the Hamiltonian we have employed the mean-field approximation to deal with the terms arising from the $\hat{S}^z\hat{S}^z$ interactions. Here, $\overline{n}_i = \langle \hat{b}_i^\dag \hat{b}_i \rangle$ is the instant occupation number of the $i$th spin site. The above Hamiltonian can be cast into the compact form of $\hat{H}(t) = \hat{\bf R}^{\dagger} V(t)\hat{\bf R}$ where we have introduced the bosonic operator vector $\hat{\bf R}=(\hat{a}, \hat{b}_{1}, \cdots, \hat{b}_{N})^{\sf T}$ and the dynamical matrix:
\begin{equation*}
V \!=\left(\begin{array}{ccccc}
\omega_0 & J_1 & \hspace{-5mm}J_2 & \hspace{-1mm}\cdots & \hspace{-1mm}J_N \\
J_1 & h+\tfrac{\lambda}{2}(\overline{n}_2 -1) & \hspace{-5mm}\lambda & \hspace{-1mm}\cdots & \hspace{-1mm}0 \\
J_2 & \lambda & \hspace{-5mm}h+\tfrac{\lambda}{2}(\overline{n}_1+\overline{n}_3 -2) & \hspace{-1mm}\cdots  & \hspace{-1mm}0 \\
\vdots & \vdots & \hspace{-5mm}\vdots & \hspace{-1mm}\ddots & \hspace{-1mm}\vdots \\
J_N & 0 & \hspace{-5mm}0 & \hspace{-1mm}\cdots & \hspace{-1mm}h+\tfrac{\lambda}{2}(\overline{n}_{N-1} -1)
\end{array}\right).
\end{equation*}
This is an $(N + 1)\times (N + 1)$ matrix, where $N$ is the total number of spins in the bath.
\begin{figure}[tb]
\includegraphics[width=\columnwidth]{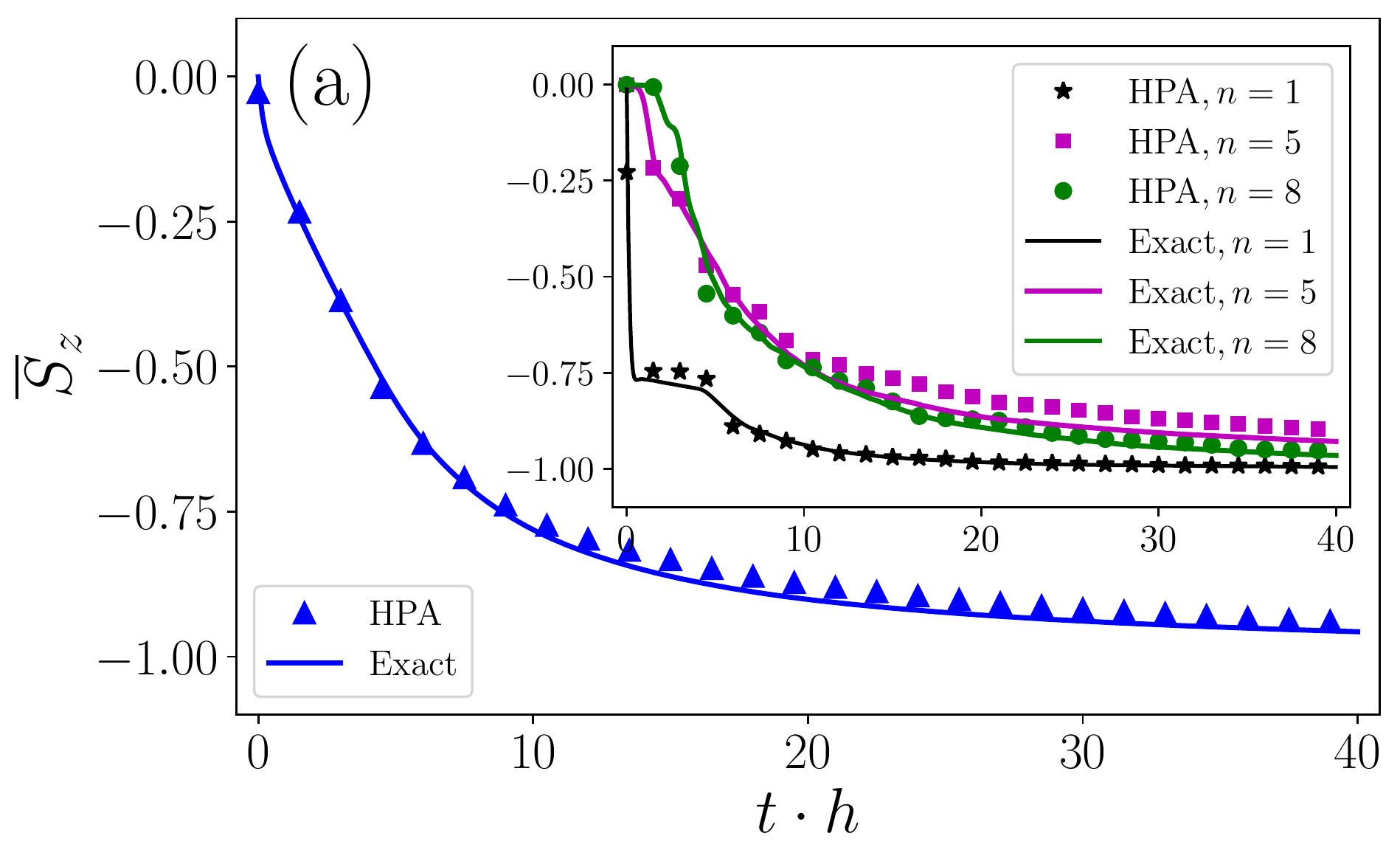}\\
\includegraphics[width=\columnwidth]{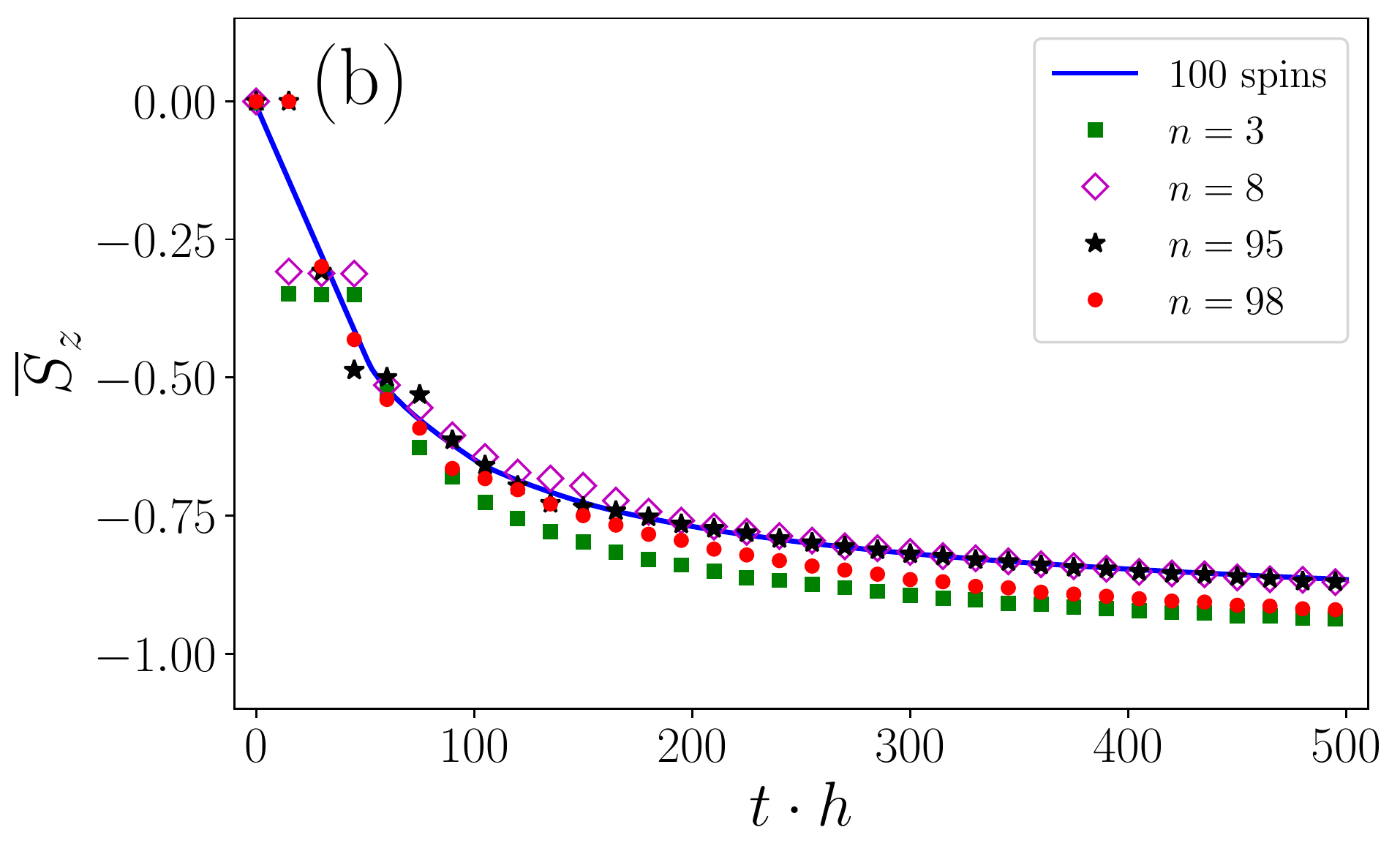}
\caption{%
The hyperpolarization dynamic of the Heisenberg model, $\overline{S}_{\!z}$, as a function of the  normalized polarization time  $t\cdot h$. (a) Comparison of the exact numerical simulation and  HPA method for $N=9$ spins with the polarization cycle time $\tau=5/h$ and for $R=800$. The inset in (a) displays the polarization dynamics of the individual spins.
(b) The  polarization dynamics for $N=100$ based on the  HPA approach with $R=10000$. The parameters are $\omega_0=h=100 $, $ \lambda=2$, $ J=10$, and $ \alpha=2 $.
}
\label{Fig1}
\end{figure}

Since the initial spin bath state is assumed to be thermal and the control spin is put in the ground state, thus for the initial state one has $\mean{\hat{\bf R}}=0$ and it retains this value given the purely quadratic  form of Eq.~\eqref{i9}.
Starting from the von Neumann equation, it is straightforward to show that the evolution of the covariance matrix, $\Gamma_{i,j} = \langle \hat{R}_i^\dag \hat{R}_j\rangle$, is given by the following
\begin{equation}
\dot \Gamma = - i \,[V(t), \Gamma] \text{ , }
\end{equation}
whose formal solution is
\begin{equation}
\Gamma(t) =\mathcal{U}(t)\,\Gamma(0) \,\mathcal{U^\dag}(t)\text{ , }
\label{i10}
\end{equation}
in which $\mathcal{U}(t)=\exp\left[ -i\int_0^t V(s)ds\right] $.

Furthermore, Eq.~(\ref{i6}) can be expressed based on the bosonic operators 
\begin{align}
\overline{S}_{\!z}&=\frac{1}{N}\sum_{n=1}^N\frac{ \mean{\hat{b}_n^\dag \hat{b}_n}-s_n}{s_{n}}\text{ . }
\label{i7}
\end{align}
We evaluate the hyperpolarization process by investigating the time evolution of $\overline{S}_{\!z}$ when the total initial state is set to $\rho(0)=\proj{0}_{\rm cs}\bigotimes^{N}_{n=1}\rho^{\rm th}_{n}$. 
Henceforth, we take $J_n=J/n^{\alpha}$ for the coupling of the control spin to the Heisenberg chain, without loss of generality:
For $\alpha = 0$ the coupling is homogeneous, while other values of $\alpha$ represent a long range coupling with various scalings.
As $\alpha \to \infty$ one retrieves the nearest neighbor coupling of the control spin to the spin at site $n=1$.
\begin{figure*}[tb]
\includegraphics[width=1.2\columnwidth]{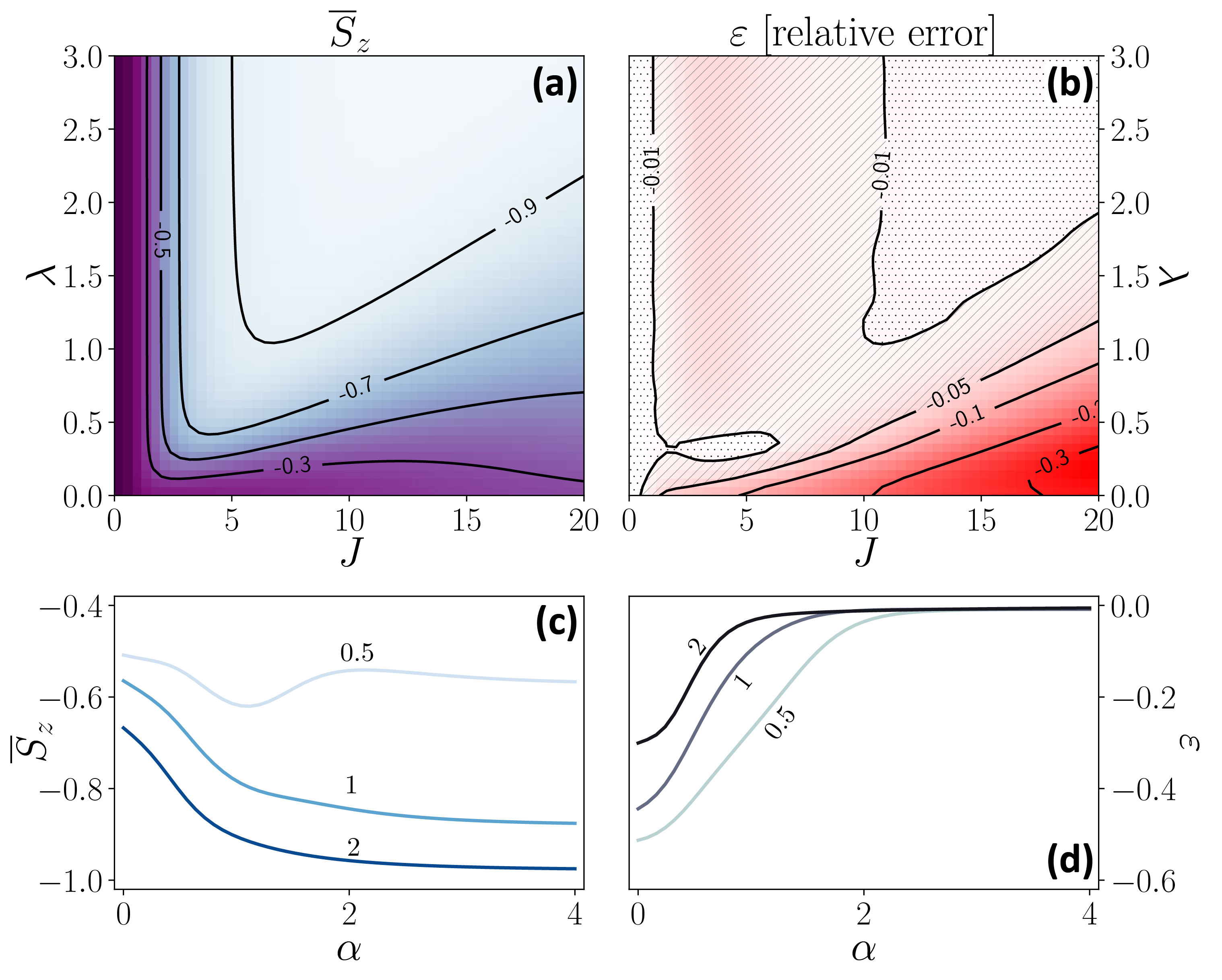}
\caption{%
The exact numerical simulation of hyperpolarization in the central spin model and its comparison to the Gaussian method:
(a) The exact polarization and (b) its relative with respect to the results from the Gaussian method $\varepsilon$ as a function of $J$ and $\lambda$ for $\alpha=2$.
(c) Variations of the polarization and (d) the difference with respect to the coupling exponent $\alpha$ when $J=10$.
The numbers next to each line correspond to a different value of $\lambda$.
The other parameters are $\omega_0=h=100 $, $ \tau=5/h$, $N=7$, and $R=500$.
}
\label{Fig2}
\end{figure*}
We first perform the numerical analysis with both exact and the approximate methods for $N=9$. 
The results are shown in Fig.~\ref{Fig1}(a).
We compare the numerical outcomes of equations \eqref{i6} and \eqref{i7} that are referred as exact and bosonic (HPA) approaches, respectively.
In generating the plot we employ the following parameters $\omega_0=h=100$, $\lambda=2$, $ J=10$ and $\alpha=2$.
The polarization duration of each cycle is numerically optimized to $ \tau=5/h$. In fact, several different values of $\tau$ have been inspected, and the best value has been chosen for the greatest polarization.
As can be seen, the exact (solid line) and the HPA (triangles) are well matched and approach to $\overline{S}_{\!z}=-1$.
It is insightful to know how polarization of the individual spins evolves.
Hence, we interrogate $\mean{S^{z}_n}/s_{n}$ for $n = 1,5,8$ and plot them in the inset of Fig.~\ref{Fig1}(a).
The first spin in the chain (black line and stars) gets polarized quickly, as one would expect.
Surprisingly, the fifth spin (magenta line and squares) is polarized slower than the eighth spin (green line and circles).
We shall immediately discuss about this behavior.
The agreement between the exact curves and HPA solutions implies that the HPA is valid in this system for the chosen set of parameters.
Therefore, this approximate method is applicable for larger spin systems where the exact solution is not possible due to the necessity of enormous computation resources.

Next, we examine the polarization behavior for $N=100$ based on the bosonic solution for the same system parameters.
The solid blue line in Fig.~\ref{Fig1}(b) shows that the total polarization can reach to values as high as $\overline{S}_{\!z}\approx -0.8$ after $R=10000$ realizations. Higher amount of polarization is attainable for larger number of cycles, and thus, longer times.
Moreover, while inspecting the individual spin polarizations we observe that the polarization of the spins at the beginning ($n=1,\cdots,7$) and at the end ($n=96,\cdots,100$) of the chain attain a larger amount of polarization  compared to the rest of the spins.
Meanwhile, the rate of polarizing is higher in the former group.
This can be understood by considering the fact that these two groups have fewer number of neighbors from one of their ends, and hence, their affection from the chain is smaller.
Yet, the first group ($n=1,2,\cdots,7$) is closer to the control spin and polarizes faster.
These can be seen from Fig.~\ref{Fig1}(b) where the polarizations of the third (green squares) and the $98$th spin (red circles) are plotted as the representatives of the first and second groups, respectively.
Moreover, the polarization of $8$th (magenta diamonds) and $95$th (black stars) spins are also shown revealing behavior of the `intermediate' spins.

We now analyze the efficiency of the hyperpolarization as well as the accuracy of the approximate method employed in this work for various system parameters.
In Fig.~\ref{Fig2}(a), we demonstrate the effect of spin interaction on hyperpolarization by using exact numerical simulation for $N=7$ and $\alpha=2$.
As can be observed, when $\lambda$ is zero, that means there is no  interaction among spins, the hyperpolarization is negligible even for large $J$ values.
By increasing $\lambda$ the polarization is improved and gets close to $\overline{S}_{\!z}\approx -1$ for strong enough coupling rates to the central spin ($J>5$).
These, show that the interaction among the bath spins works constructively for the sake of hyperpolarization throughout the one-dimensional lattice.
A similar study is performed on the effect of the long range interaction exponent, and reported in Fig.~\ref{Fig2}(c), where $\overline{S}_{\!z}$ is plotted against $\alpha$ for three different values of $\lambda$.
For large enough inter-spin coupling rates larger values of $\alpha$ will favor the polarization.
We compare these exact numerical results with the corresponding bosonic method by defining the relative error as $\varepsilon = \big([\overline{S}_{\!z}]_{\rm exact} -[\overline{S}_{\!z}]_{\rm HPA}\big)/\big|[\overline{S}_{\!z}]_{\rm exact} +[\overline{S}_{\!z}]_{\rm HPA}\big|$\gtxt{,} which assumes both positive (overestimating the exact results) and negative (underestimating the exact results) values.
The relative error behavior is shown in Fig.~\ref{Fig2}(b) where the dotted area implies the difference between the exact and the bosonic methods negligible and less than $\%1$, while the hashed area corresponds to the parameter region where the HPA method underestimates the exact results with less than $\%5$ error.
Note that our goal is to find the parameters that both give a high polarization and low relative error.
Hence, the highest polarization associated with the least relative error can be generally obtained for $J \geq 10$ and $\lambda\geq 2$.
These two optimal limits also explain our choice of parameters in Fig.~\ref{Fig1}.
A similar study is performed for $\alpha$ and it is found that the relative error saturates and becomes less that one percent for $\alpha\geq 2$, see Fig.~\ref{Fig2}(d).

By this introduction to the possibility of hyperpolarization by an optically active spin in a simple one-dimension lattice and the  validity of bosonic model, in the next section we consider the same problem for a real two-dimensional hexagonal Boron Nitride (hBN) lattice with a defect.
%
\section{hexagonal boron nitride lattice} \label{hBN}%
Hexagonal boron nitride is an ideal van der Waals crystal for hosting optically active defects since it has a large bandgap ($\approx 6$~eV)~\cite{Cassabois2016, Tran2016}.
Defects in hBN can be exploited for flexible two-dimensional quantum sensors~\cite{healey2021q, Tetienne2021b}.
Furthermore, there are proposals for employing the freestanding layers of hBN as optomechanical systems~\cite{Abdi2017, Abdi2019}, which are based on the exceptional mechanical properties of the hBN and the accompanying color centers~\cite{Song2010}.

Each site in the hBN lattice is occupied by a nitrogen nuclear or a boron nuclear that they have a non-zero spin. A schematic of the hBN lattice with a \vnnb\ defect is shown in Fig.~\ref{Fig3}(b) which consists of a missing boron atom.
The negatively charged boron vacancy center has a triplet spin ground state ($S=1$) with zero-field splitting $D/2\pi\approx 3.5$~GHz.
The simplified energy level diagram of the \vnnb\ defect is depicted in Fig.~\ref{Fig3}(c) that its electronic ground state triplet can be efficiently polarized using optical excitation~\cite{Gottscholl2020}.
Hence, this defect can be fully controlled through microwave and laser drives.
In fact, the defect radiates a red light at a wavelength centered at $\approx 850$~nm when excited with a green $532$~nm laser.
In the following sections we introduce the nuclear spin bath model and then study the possibility of hyperpolarizing the hBN spin lattice.
For the sake of simplicity, we shall assume that the sample is purified to the ${}^{11}\mathrm{B}$ isotopes.
%
\subsection{The nuclear spin bath model}\label{suba}%
The hyper-fine (hf) interaction with two mechanisms and their different physical origins can contribute to the coupling between an electron spin $\mathbf{S}$ and a nuclear spin $\mathbf{I}$.
The first mechanism is the dipole-dipole interaction between the magnetic moments of the electron and nuclear spins.
In analogy to the classical dipolar interaction between magnetic moments its Hamiltonian is written as
\begin{equation}
\hat H_{\rm dd}^{e,i} = \frac{\mu_0}{4\pi}\frac{\gamma_e\gamma_{n,i}}{r_i^3}\Big[\hat{\bf S}\cdot\hat{\bf I}_{i} -\frac{3(\hat{\bf S}\cdot\mathbf{r}_i)(\hat{\bf I}_{i}\cdot\mathbf{r}_i)}{r_i^2}\Big],
\label{hamdd}
\end{equation}
where $\mu_0$ is the permeability of the vacuum and $\mathbf{r}_i$ is the displacement vector pointing from electron to the $i$th nuclear (in meters) and $\hat{\bm r}_{i}$ is the   corresponding unit vector. Here, $\gamma_e$ and $\gamma_{n,i}$ are the gyromagnetic ratios of electron and nucleus at site $i$, respectively.
The dipole-dipole interaction depends on the relative orientation of the magnetic moments and is thus anisotropic.
Purely dipolar interaction is expected if the electron spin is located in a molecular orbital with no overlap with the nuclei.
In contrast, the second hf mechanism becomes important if there is a finite probability of the electron presence at the location of the nuclei.
In fact, this is the case only if there are contributions of $1s$-orbital (totally symmetric orbital) to the molecular orbitals that accommodate the electron of interest~\cite{Langhoff2012, Doherty2013}.
The energy term of this so-called Fermi contact interaction is given by
\begin{equation}
\hat H_{\rm Fc} = \frac{8\pi}{3}\frac{\mu_0}{4\pi}\gamma_e\sum_i\gamma_{n,i}\abs{\Psi(\mathbf{r}_i)}^2\hat{\bf S}\cdot\hat{\bf I}_{i}.
\label{hamfc}
\end{equation}
Here, $\abs{\Psi(\mathbf{r}_i)}^2$ is the probability density of the electron in the orbital described by the wave function $\Psi$.
Evidently, the Fermi contact interaction is isotropic.

In the \vnnb\ defect the electron occupies the well-localized $\pi$-type orbitals, which has no overlap with the lattice nuclei~\cite{Abdi2018}.
Hence, only the dipole-dipole term (\ref{hamdd}) in the hf interaction has significant contribution.

We now study the hyperpolarization of a mono-layer of hexagonal boron nitride  lattice by the electron spin of \vnnb~defect.
We consider the nuclear spin lattice of a mono-layer of hBN of arbitrary size.
We assume an isotope purified sample and only take into account the naturally prominent isotope of boron, that is $^{11}$B, which naturally take 80 percent proportion of the total.
The nuclear spins of the different constituents of the lattice are the following
\begin{equation}
S_{\rm V_B} = 1, \qquad S_{\rm N} = 1, \qquad S_{\rm B} = 3/2\text{.}
\end{equation}

Here, we suggest a protocol that can acquire rapid hyperpolarization of both nitrogen and boron nuclear spins at room temperature, where those spins are fully thermal. 
The electron spin of the \vnnb\ defect is first polarized with a short laser pulse.
Then the polarization is transferred to the nuclei by means of a microwave drive and thanks to its hyperfine interaction with the nuclei.
This, nonetheless, is optimal when the Hartmann-Hahn resonance is satisfied.
The process is repeated by polarizing the electron spin and continues as many times as necessary to arrive at the highest polarization level.
Note that even in ideal conditions one cycle of the polarization transfer is not enough for polarizing one single nuclear spin (boron or nitrogen).
Because the spin of nuclear spins is higher than that of the effective one-half spin of the defect, see below.
Therefore, to obtain a strong polarization, it is requisite to exploit many reiterations of the polarization protocol.

Meanwhile, the nuclear spins are coupled to each other via the dipole-dipole interactions:
\begin{equation}
\label{eq:dipHam}
\hat{H}_{\rm dd}^{ij} = \hbar g_{ij} \big[ \hat{\bf I}_i\cdot\hat{\bf I}_j - 3 (\hat{\bf I}_i\cdot \hat{\bm r}_{ij})(\hat{\bf I}_j\cdot \hat{\bm r}_{ij}) \big]\text{,}
\end{equation}
where $g_{ij} = \hbar \mu_0 \gamma_{n,i} \gamma_{n,j}/4 \pi |\mathbf{r}_{ij}|^3$ is the coupling strength and $\hat{\bf I}_i = (\hat{I}_i^x, \hat{I}_i^y, \hat{I}_i^z)$ is the vector of spin operators.
Here, $\mathbf{r}_{ij}=\mathbf{r}_i-\mathbf{r}_j$ is the spatial vector connecting the two interacting spins, with $\hat{\bm r}_{ij} = (\hat{x}_{ij},\hat{y}_{ij},\hat{z}_{ij})$ its corresponding unit vector.
The lattice constant in hBN is $\approx 1.5$~\AA, while the boron and nitrogen gyromagnetic ratios are $\gamma_{\rm B}/2 \pi = 13.66$~MHz/T and $\gamma_{\rm N}/2\pi = 3.078$~MHz/T, respectively.
As it shall become clear in the next section, unlike the simple one-dimensional Heisenberg chain these inter-nuclei interactions do not contribute in the hyperpolarization process, in contrast, they can be restrictive.

%
\subsection{Engineering the system dynamics}\label{subb}
We take into account $N$ symmetrically nearest nuclear spins to the defect and assume a high external magnetic field such that $D \ll \gamma_e|\mathbf{B}|$,  see the inset in Fig.~\ref{Fig3}(c).
Therefore, the defect natural quantization axis is suppressed and the magnetic field
orientation dictates the quantization axis for both electron and nuclei, which we assign to the  $z$-axis. 
By applying a microwave driving field with Rabi frequency $\Omega$ and focusing on the $\left\lbrace \ket{-1}, \ket{0}\right\rbrace $ defect subspace as well as under rotating wave approximation,
the total effective Hamiltonian describing dynamics of the system  in the rotating frame with the driving field frequency $\omega_{\rm mw}$ reads
\begin{align} \label{Heff}
\hat H_{\rm eff} &= \Omega \hat{s}^{z} +\sum^{N}_{i=1} \acute{\omega}_{i} \hat{I}^{z}_{i} +\sum^{N}_{i=1} (\alpha_{i} \hat{s}^{+}\hat{I}^{-}_{i} +\alpha^{\ast}_{i} \hat{s}^{-}\hat{I}^{+}_{i})\nonumber\\
&+\sum^{N}_{i< j} b_{ij}\left( \hat{I}_i^z \hat{I}_j^z -\frac{1}{4}(\hat{I}_i^+ \hat{I}_j^- + \hat{I}_i^- \hat{I}_j^+)\right) \text{ , }
\end{align}
where $\hat{s}^{z}=\frac{1}{2}(\trans{+}{+} -\trans{-}{-})$ and $ \hat{s}^+=\trans{+}{-}= (\hat{s}^-)^{\dag}$ are the Pauli matrices in the dressed state basis $\ket{\pm} = (\ket{0}\pm\ket{-1})/\sqrt{2}$ representation, see the Appendix for details of the derivation.
Here, the hyperfine coupling $\alpha_{i}=\frac{1}{4} (A^{x}_{i}+i\,A^{y}_{i})$, the modified nuclear Larmor frequencies $\acute{\omega} _{i}=\omega_{i}-\frac{1}{2}A^{z}_{i}$, and the inter-nuclei dipolar coupling rate $b_{ij} = g_{ij}(1-3\hat{z}_{ij}^{2})$ have been introduced.
The hyperfine coupling vector of the electron spin to the $i$th nuclear spin is given by $\mathbf{A}_{i}=g_{ei} (-3\hat{x}_i\hat{z}_i, -3\hat{y}_i\hat{z}_i, 1-3\hat{z}_i^{2})$ with $g_{ei} = \hbar \mu_0 \gamma_{e} \gamma_{n,i}/4 \pi |\mathbf{r}_{i}|^3$.

It is clear from the above Hamiltonian and the hyperfine vector that hyperpolarization of a monolayer of hBN is not possible when the quantization axis is perpendicular to the layer.
Because in this case $\alpha_{i}=0$, and thus, the flip-flop interactions between the defect and the nuclear spins cannot occur. To circumvent this, we assume that a high magnetic field, tilted with respect to the layer, is applied with the deviation angle $\theta$, see Fig.~\ref{Fig3}(b) for the sketch. Note that in the limit of weak magnetic field the natural quantization axis of \vnnb\ in its D$_{3h}$ symmetry is perpendicular to the layer, and thus, the hyperpolarization is inaccessible. This also explains our choice of large magnetic field at the beginning of this section.
 
We next find the angle $\theta$ that maximizes the flip-flop interactions responsible for the polarization transfer.
It is straightforward to obtain 
\begin{equation}
\vert \alpha_{i}\vert = \frac{1}{4}\sqrt{\vert A^{x}_{i}\vert^{2}+\vert A^{y}_{i}\vert^{2}}=\frac{3}{8}g_{ei} |\sin (2\theta)|\text{ .}
\end{equation}
Therefore, the greatest amount of $\vert \alpha_{i}\vert$ is found for $\theta=\frac{\pi}{4}$.
Nonetheless, the azimuthal angle of the magnetic field also plays an important role here and can be employed for tuning the coupling of the defect to different sites.
Due to tremendous complexity, we find the optimal azimuthal angle numerically.

%
\subsection{Hyperpolarization of hBN}\label{subc}
The hyperpolarization scheme is similar to the one explained in Sec.~\ref{sec:isinghyper} and can be summarized as follows:
First, by using an optical pumping the electron spin of the ground state of the \vnnb\ defect is brought to the $\ket{0}$, i.e. the state with $m_{s}=0$.
Then it is prepared in the superposition state $\ket{-} = (\ket{0}-\ket{-1})/\sqrt{2}$ by applying a $\tfrac{\pi}{2}$-pulse~\cite{Gottscholl2020}. 
Second, the polarization is transferred from the \vnnb\ spin to the nuclear spins via the hyperfine interactions and by applying a resonant microwave drive.
The steps of the polarization cycle are iterated multiple times.
Nonetheless, one should keep in mind that since there are two spin species (three for the natural isotope shares) in the lattice, the resonance condition cannot simultaneously hold for all the bath spins.
Given the long coherence time of the nuclear spins this can be circumvented, e.g. by alternating the resonant condition for all spin species of interest.
The whole initial state is $\rho(0)=\rho_{e}(0)\bigotimes^{N}_{i=1} \rho^{\rm th}_{i}$ where the initial state of the defect is set to $\rho_{e}(0)=\proj{-}$.
Meanwhile, the nuclear spins are assumed to be in thermal equilibrium with independent reservoirs, and thus, their initial density matrix is given by $\rho^{\rm th}_{i}$.
The evolution of the nuclear-spin density matrix after the $(R+1)$th cycle of the polarization protocol is determined by
\begin{align}
\hat{\rho}_{\rm nuc}\big((R+1)\tau\big) =\tr{e}{e^{-i\hat H_{\rm eff}\tau} \right(  \proj{-}_{e}\otimes\hat{\rho}_{\rm nuc}(R\tau)\left)  e^{i\hat H_{\rm eff} \tau}} \text{, }
\label{i33}
\end{align}
where $\hat H_{\rm eff}$ is the Hamiltonian \eqref{Heff}, $\tau$ is the equal duration of each cycle, and $\tr{e}{}$ denotes the trace over the electron spin.

In order to numerically simulate a large spin system we shall employ the method based on HPA explained and examined in the previous section.
Under this transformation, i.e. Eq.~\eqref{HP}, and for resonant driving of the electron ($\Delta=0$) the Hamiltonian \eqref{Heff} takes the following form
\begin{align} \label{a32}
\hat{H} (t)&=\Omega \: \hat{a}^\dag\hat{a}+\sum^{N}_{i=1} \tilde{\omega}_{i}(t)\: \hat{b}^\dag_{i} \hat{b}_{i} +\sum^{N}_{i=1} (\beta_{i}\hat{a}^{\dag}\hat{b}_{i} +\beta^{\ast}_{i}\hat{a}~\hat{b}^{\dag}_{i}) \nonumber\\
&-\sum^{N}_{{\substack{i< j}}} B_{ij}(\hat{b}^{\dag}_{i}\hat{b}_{j} +\hat{b}_{i}\hat{b}^{\dag}_{j}) \text{ , }
\end{align}
where we have introduced the electron-nuclei coupling rate $\beta_{i}= \sqrt{2s_{i}}\alpha_{i}$ and nuclear-nuclear interaction strengths $B_{ij}=\frac{1}{2}\sqrt{s_{i}s_{j}}\: b_{ij}$.
Here, the modified Larmor frequency of the nuclear spins is $\tilde{\omega}_{i}(t)=\acute\omega_{i} +\half\sum_{j\neq i}  b_{ij}\left(\overline{n}_j(t) -2s_{j}\right)$ where the time-dependent expectation value is due to the mean-field approximation.
This Hamiltonian is quadratic in the bosonic operators, and this allows for an efficient and tractable numerical simulation describing the dynamical evolution via the covariance matrix formalism.
The above Hamiltonian according to the bosonic operator vector $\hat{\bf R}=( \hat{a}, \hat{b}_{1}, \cdots,  \hat{b}_{N})^{\sf T}$  can be expressed as $\hat{H}(t)=\hat{\bf R}^{\dagger} W(t)\hat{\bf R}$, with the dynamics matrix
\begin{equation}
W(t)=\left(\begin{array}{cccccc}
\Omega &\beta_{1} & \beta_{2}& \beta_{3}&\cdots &\beta_{N}\\
\beta^{\ast}_{1} & \tilde{\omega}_{1}(t) &- B_{12}&-B_{13}  & \cdots &-B_{1N}\\
\beta^{\ast}_{2} & -B_{12}& \tilde{\omega}_{2}(t) &-B_{23} & \cdots  & -B_{2N} \\
\beta^{\ast}_{3}&-B_{13} &- B_{23} & \tilde{\omega}_{3}(t)& \cdots  &-B_{3N}\\
\vdots & \vdots & \vdots & \vdots & \ddots & \vdots \\
\beta^{\ast}_{N} &- B_{1N}& -B_{2N}&-B_{3N} & \cdots & \tilde{\omega}_{N}(t) \\
\end{array}\right)\text{ . }\\\nonumber
\end{equation}
Thanks to the quadratic form of the Hamiltonian \eqref{a32} the system is fully characterized by the first and the second moments of the operators $\hat{\bf R}$.
The covariance matrix, $\Gamma_{i,j} = \langle \hat{R}_i^\dag \hat{R}_j\rangle$, is sufficient for describing the system dynamics when the initial state gives zero-mean value.
In our case, the nuclei are initially set to a thermal state and the electron is in the ground state of the dressed state basis.
Therefore, we have  $\langle \hat{\bf R}\rangle=0$ and proceed with analyzing the covariance matrix dynamics by employing Eq.~\eqref{i10}.

\begin{figure}[tb]
\includegraphics[width=\columnwidth]{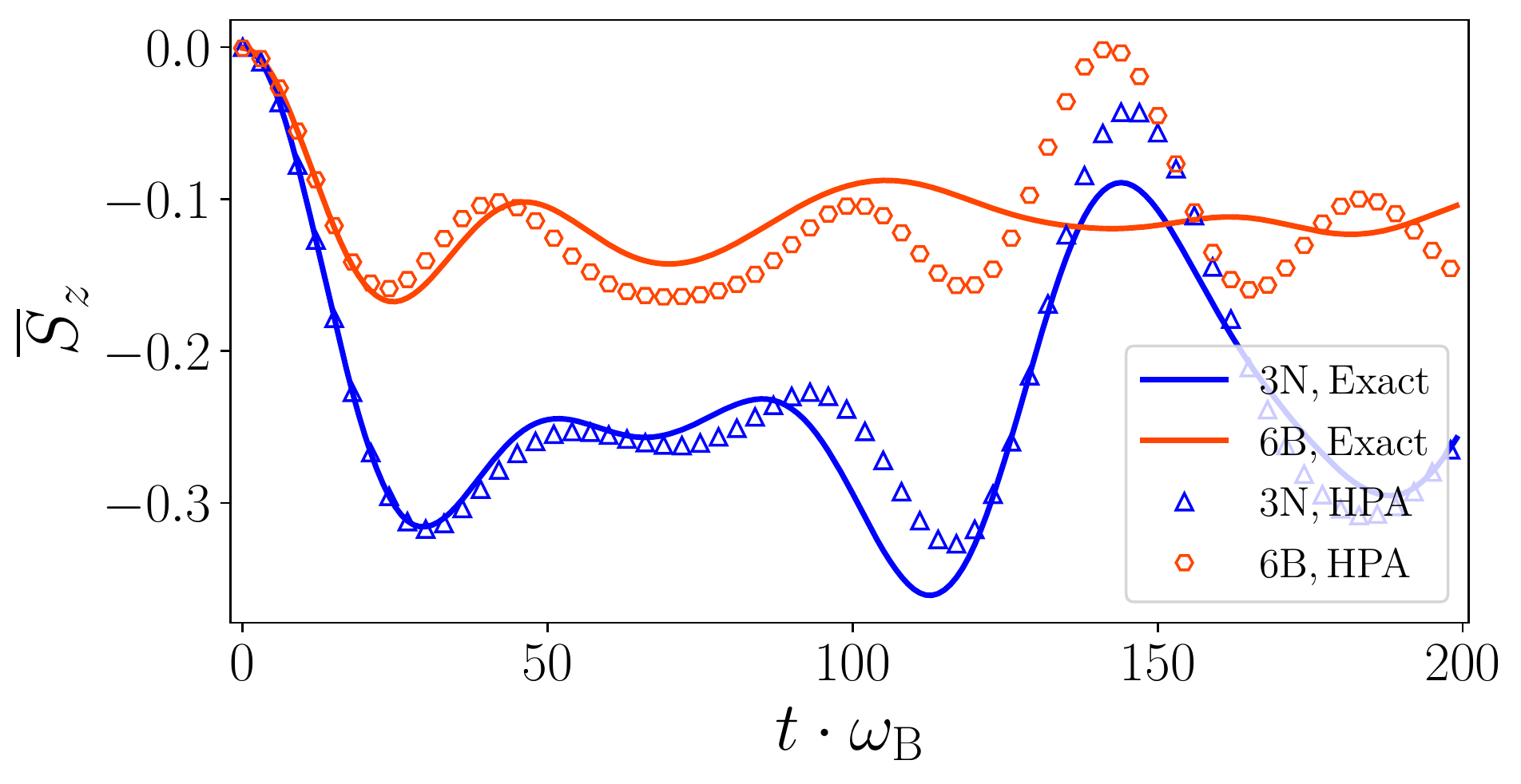}\\
\caption{%
The short time polarization dynamics of the defect and the nearest nuclei:
$\overline{S}_{\!z}$ is given separately for the three nitrogens (blue) and six borons (red) who are the nearest nuclei to the \vnnb.
The exact results are shown by solid lines, while the HPA method are given as markers.
The \vnnb\ and the nuclei states are initially set to $\ket{-}$ and a fully thermal state, respectively. The magnetic field is $B=1$~T and the Rabi frequencies are set to satisfy the Hartmann-Hahn resonance conditions.}
\label{Fig4}
\end{figure}
Before studying large-size systems, we compare the HPA technique and the exact numerical simulation for first two rings: three nearest nitrogen nuclei and six next nearest boron nuclei, see Fig.~\ref{Fig3}(b) for a graphical illustration.
To this end, we investigate the short time dynamics.
In Fig.~\ref{Fig4} the results are compared when the nuclei are initially set to the thermal state while the defect electron spin is initialized in the dressed state of $\ket{-}$.
We notice that the HPA method captures most features of the exact numerics despite the thermal initial state of the nuclei.
This is such that the results even show perfect match for $t\lesssim 25/\omega_{\rm B}$.
The deviations grow larger as the time elapses and the HPA method become less reliable.
Therefore, in our following hyperpolarization study for large systems which is based on the HPA method we always choose $\tau<25/\omega_{\rm B}$ to ensure the validity of the results.

In the hyperpolarization procedure, we take the combination of two different Rabi frequencies that each satisfy the Hartmann-Hahn resonance condition for the boron and nitrogen nuclei.
This is to guarantee the polarization of both sub-lattices.
In other words, we set $\Omega=\omega_{\rm N}$ and $\Omega=\omega_{\rm B}$ alternating in the polarization cycle iterations.
The duration of polarization for each step is numerically optimized to $\tau_{\rm N}=25/\omega_{\rm B}$ and $\tau_{\rm B}=15/\omega_{\rm B}$.
That is, many different values of $\tau_{\rm N}$ and $\tau_{\rm B}$ have been investigated and have found these the best values who give the highest polarization in the shortest time.

\begin{figure}[tb]
\centering
\includegraphics[width=\columnwidth]{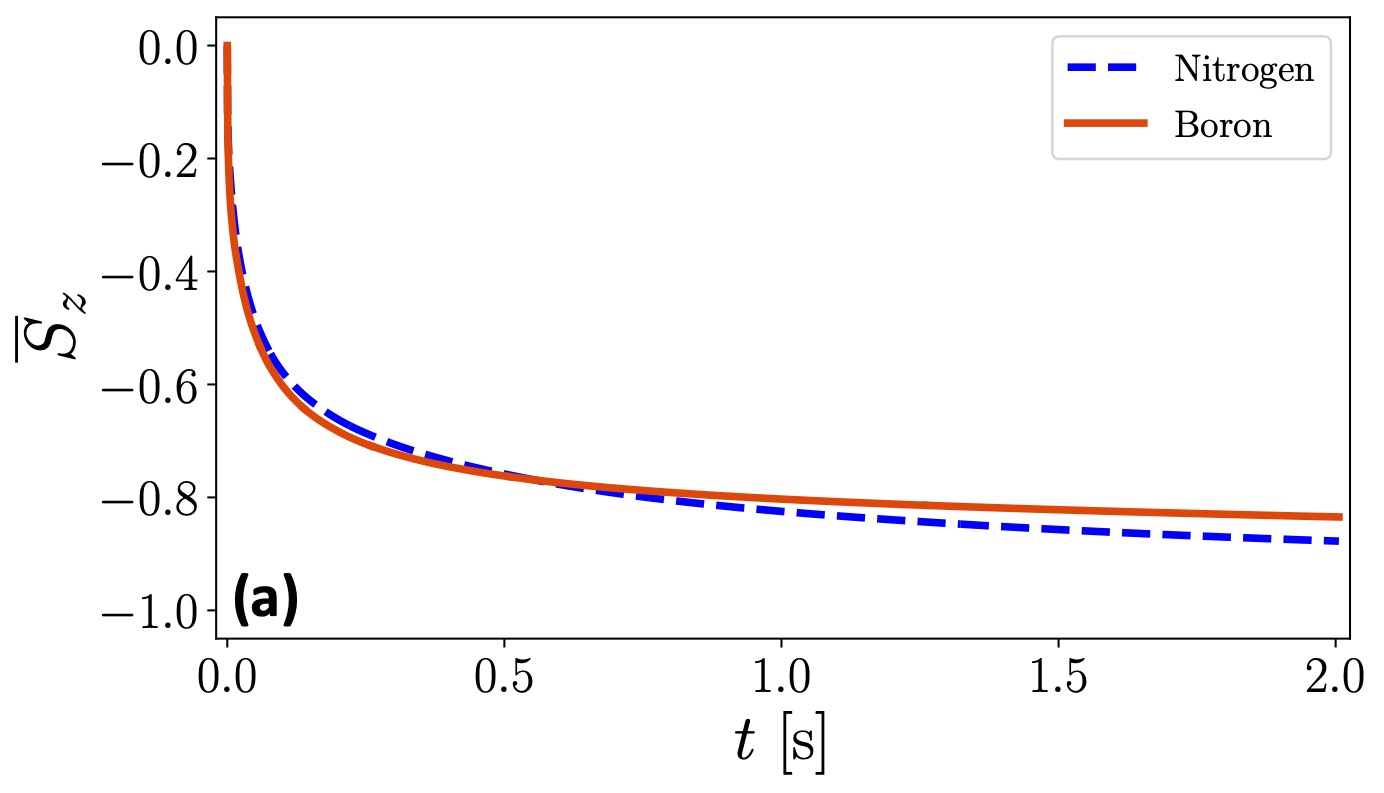}\\
\includegraphics[width=\columnwidth]{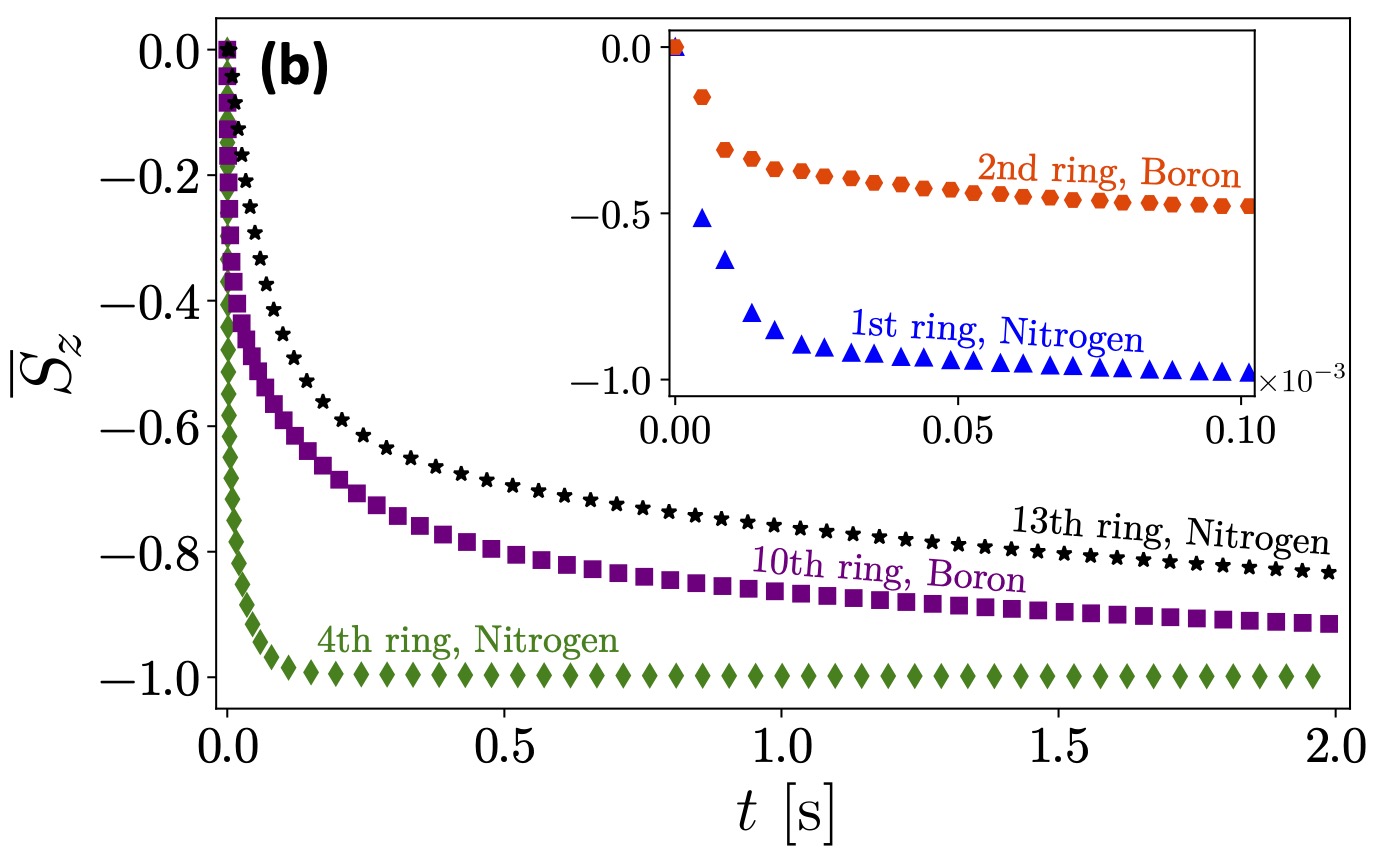}
\caption{%
Hyperpolarization of hBN lattice:
(a) The polarization dynamics is represented for 121 nuclei spins composed of 61 nitrogens and 60 borons with $R=15\times 10^{5}$.
(b) The polarization behavior for five different rings.
The inset gives the polarization evolution of the two first rings after $R=70$.
The other parameters are $\tau_{\rm N}=25/\omega_{\rm N}$, $\tau_{\rm B}=15/\omega_{\rm B}$, and $B=1$~T.}%
\label{Fig5}%
\end{figure}

Now, we extent our analysis based on the HPA method to the largest spin system that our computational resources can afford.
Hence, we assume a spin lattice with $N=121$ composed on $61$ nitrogen and $60$ boron spins.
The results are provided in Fig.~\ref{Fig5}(a), where one observes a smooth trend in the average polarization of both boron and nitrogen sub-lattices.
Nevertheless, there is a slight difference in their behavior.
The borons polarize slightly faster, due to their stronger interaction with the defect electron spin.
But their polarization already saturates after $R=15\times 10^5$ epochs.
We attribute this to the larger coupling rate among the boron nuclei, which in turn stems from their larger gyromagnetic ratio.
On the other hand, the nitrogen sub-lattice polarizes with a slower pace, yet it has the potential of approaching higher values of polarization in longer times.
Our findings suggest that---unlike one-dimensional case---in two-dimensional (and presumably higher-dimensional) lattices the strong interaction among the bath spins is prohibitive regarding the polarization transfer, see the discussion at the end of this section.

In order to better understand dynamics of the system we also examine the polarization process of the nuclei in equal distances from \vnnb\ and refer to them as the `rings' around the defects.
This is such that the 3 nearest neighbor nitrogens lie in the first ring, the second ring is composed of 6 borons in the next nearest neighbor, and so on as follows: 3 nitrogens in the third ring, 6 nitrogens in the fourth ring, 6 borons in the fifth ring.
Fig.~\ref{Fig5}(b) shows the polarization for some of the rings.
The first and second rings already get polarized to their asymptotic value after $R\approx 70$.
This rapid polarization of these two rings is understandable because of their proximity to the defect.
For the three nitrogen nuclei the weak interaction among themselves allows for a perfect polarization.
Meanwhile, the fast but limited polarization ($\overline{S}_{\!z}\approx -0.5$) of the second ring is ascribed to the competition between the inter-nuclei and defect-nuclei interactions.
However, for rings farther from the defect, the distance among borons increases and consequently the inter-nuclei interactions become much weaker.
Hence, the boron spins in the tenth ring continue their journey to the ground state, though very slowly, see the purple squares in leftmost part of the plot.
The trend is more or less the same for farther rings of both species. Note that the tenth and thirteenth rings have 12 borons and 12 nitrogens, respectively. 

To further assess the effect of inter-nuclei interactions on the polarization transfer from the \vnnb\ defect to the nuclei spins, we \textit{artificially} turn off the defect-nuclei interactions in three different scenarios: 
We first suppose that only the nitrogen spins interact with the defect and the borons are decoupled.
However, we allow them to interact among each other as well as with the nitrogen spins.
Next, we perform the same study by swapping the role of boron and nitrogen.
In another scenario we allow the defect to only interact with the first and second ring nuclei, while the rest of lattice does not have any direct interaction with the defect.
In all these investigations we generally find that the interaction between the nuclei is not able to transfer the polarization from the one specie to another or even from closer rings to the farther nuclei. 
Therefore, the weak dipole-dipole coupling of the nuclear spins, compared to the defect-nuclear interactions, the inter-nuclei interactions is favor the polarization dynamics.
Nonetheless, they can have prohibitive effects as we discussed above.

%
%
\section{Conclusion} \label{seccon}
In this work, we have presented a theoretical description of efficient polarization transfer from a \vnnb\ defect to the nuclear spins in the hexagonal boron nitride lattice at room-temperature by manipulating an optically active electron spin.
Since a large spin system can not be simulated by exact methods, we have studied the hyperpolarization via a numerical method based on the Holstein-Primakoff approximation.
By considering the central spin model as a well-understood system, we have benchmarked the method and have found that the bosonic numerical approach is in good agreement with exact numerical simulation and yields satisfactory results.

Our results can pave the way for the utilization of defects in hexagonal boron nitride for quantum technologies, high-resolution two-dimensional quantum sensors such as NMR.

\begin{acknowledgements}
FTT and MF acknowledge the support by Iran Science Elites Federation via a postdoctoral fellowship. The work of JSP and MBP was supported by ERC HyperQ. 
\end{acknowledgements}
%
%
\appendix*
\section{Derivation of the effective Hamiltonian~\eqref{Heff}} 
In this Appendix we provide the detailed steps in arriving to the effective Hamiltonian Eq.~\eqref{Heff} employed for studying the hyperpolarization of hBN nuclear spin lattice.
We begin by considering an external magnetic field which defines $z$-axis of the laboratory frame of reference and assume a hBN layer that the \vnnb\ defect is located at the origin.
It should be noted that \vnnb\ has a D$_{3h}$ symmetry with the axis of symmetry perpendicular to the layer such that the misalignment angle between the magnetic field and \vnnb-axis is given by $\theta$, see Fig.~\ref{Fig3}(b).
We suppose the applied magnetic field is strong, $D \ll \gamma_e|\mathbf{B}|$, such that the quantization axis is determined by the magnetic field and the zero-field splitting will be regarded as a perturbation that leads to energy shifts of the Hamiltonian diagonal elements~\cite{Chen2015, Scheuer2016}. The total Hamiltonian describing dynamics of the system  in the laboratory frame reads $\hat H=\hat H_{0}+\hat H_{I}$,
where $\hat{H}_0$ constitutes the \vnnb\ defect and the nuclear spin lattice free Hamiltonians:
\begin{align} \label{ap1}
\hat H_{0}&=\big[\omega_e +\delta (\theta)\big] \hat{S}_{z} + D(\theta)\hat{S}_{z}^2
+\sum^{N}_{i=1} \omega_{i}\: \hat I^{z}_{i} \text{ , }
\end{align}
where $\omega_e = -\gamma_e B$ and $\omega_i=\gamma_{n,i}B$ are respectively the Larmor frequencies of electron and the nuclei, and we have introduced
\begin{align*}
D(\theta) &= \frac{D}{4} (1+3 \cos2\theta)\text{ , }\\
\delta(\theta) &= \frac{D^2}{8\omega_e}\Big[\sin^4\!\theta + \frac{\sin^2(2\theta)}{1- \big(D(\theta)/\omega_e\big)^2} \Big]\text{ , }
\end{align*}
where $D$ denotes the zero-field splitting of the defect~\cite{Scheuer2016}.

The interaction Hamiltonian is given by
 \begin{align} 
\hat H_{I}&=\sum^{N}_{i=1}\hat H^{e,i}_{\rm dd}+\sum^{N}_{{\substack{i< j}}}\hat H^{ij}_{\rm dd}\text{ . }
\end{align}
Before proceeding, we simplify the interactions by applying the secular approximation.
For this purpose we first express the interactions in terms of the spin raising and lowering operators $\hat{I}^\pm_i = \hat{I}^x_i \pm i\hat{I}^y_i$.
Hence, Eq.~\eqref{eq:dipHam} becomes
\begin{align}\label{a6}
\hat H_{\rm dd}^{ij} &= b_{ij}\big[ \hat{I}_i^z \hat{I}_j^z -\frac{1}{4}(\hat{I}_i^+ \hat{I}_j^- + \hat{I}_i^- \hat{I}_j^+)\big] \nonumber\\
&+\Big[c_{ij}(\hat{I}_i^z \hat{I}_j^+ + \hat{I}_i^+\hat{I}_j^z) +d_{ij}(\hat{I}_i^+ \hat{I}_j^+) + \text{H.c.}\Big] \text{ , }
\end{align}
where have introduced the following coefficients
\begin{align*}
b_{ij} &= g_{ij}(1-3\hat{z}_{ij}^{2}), \\
c_{ij} &= -\tfrac{3}{2}g_{ij}(\hat{x}_{ij}-i\:\hat{y}_{ij}) \hat{z}_{ij}, \\
d_{ij} &= -\tfrac{3}{4}g_{ij}(\hat{x}_{ij}^{2}-\hat{y}_{ij}^{2}-2i\hat{x}_{ij}\hat{y}_{ij}).
\end{align*}
In the interaction picture of the free nuclear spin lattice Hamiltonian, the nuclear spins transform as
\begin{equation}
\hat{I}_i^\pm \rightarrow \hat{I}_i^\pm e^{\pm i \omega_i t}, \qquad \hat{I}_i^z \rightarrow \hat{I}_i^z \text{.}
\end{equation}
Under the rotating wave approximation, which is valid if the applied magnetic field is high, and thus, the Larmor frequencies are much larger than the dipole-dipole coupling, i.e. $\lbrace\omega_i, \omega_{j} \rbrace\gg g_{ij}$, one neglects the effect of the resulting time-dependent terms in Eq.~(\ref{a6}).
Hence, reducing the spin-spin interaction to 
\begin{equation}
H_{\rm dd}^{ij} \approx b_{ij} \big[\hat{I}_i^z \hat{I}_j^z - \dfrac{1}{4}(\hat{I}_i^+ \hat{I}_j^- + \hat{I}_i^- \hat{I}_j^+) \big]\text{ . }
\end{equation}

For the interaction of the defect electron spin with the nuclear spins, which have gyromagnetic ratios differing by about 3 orders of magnitude one has
\begin{align}
\hat H^{ei}_{\rm dd}&\approx \hat{S}_{z}\:\mathbf{A}_{i}\cdot \hat{\mathbf{I}}_{i}\text{ , }
\end{align}
where  $\mathbf{A}_{i}=g_{ei} (-3\hat{x}_i\hat{z}_i, -3\hat{y}_i\hat{z}_i, 1-3\hat{z}_i^{2})$ with $g_{ei} = \hbar \mu_0 \gamma_{e} \gamma_{n,i}/4 \pi |\mathbf{r}_{i}|^3$, is the hyperfine coupling vector of the electron spin to the $i$th nuclear spin.
Note that we have simplified the hyperfine interaction by secular approximation, since the nonsecular hyperfine coupling can be neglected at magnetic fields that provide large differences between the electronic and nuclear Zeeman splitting, i.e., $\abs{\omega_e +\delta (\theta) -\omega_{i}}\gg g_{ei}$.
Hence, after applying the secular approximations the Hamiltonian describing the system dynamics reads
\begin{align} \label{a28}
\hat H &=\big[\omega_e +\delta (\theta)\big] \hat{S}_{z} + D(\theta) \hat{S}_{z}^2 +\sum^{N}_{i=1} \omega_{i} \hat{I}^{z}_{i}\nonumber\\
&+\sum^{N}_{i=1} \hat{S}_{z}\: \mathbf{A}_{i}\cdot \hat {\bf I}_{i} 
+\sum^{N}_{i< j} b_{ij}\left( \hat{I}_i^z \hat{I}_j^z -\frac{1}{4}(\hat{I}_i^+ \hat{I}_j^- + \hat{I}_i^- \hat{I}_j^+)\right)  \text{ . }
\end{align}
The flip-flop interaction of the electron spin with the nuclei---necessary for polarization transfer---is negligible as $\omega_e \gg \omega_i$.
Nevertheless, an efficient two-level system that is on resonance with the bath spins is constructed by a microwave drive with a proper frequency and amplitude.
In our analysis we assume that the microwave field applied to the \vnnb\ defect is tuned for driving the spin transition $\ket{0}\leftrightarrow\ket{-1}$.
The corresponding Hamiltonian is then
\begin{align}
\hat H_{\rm mw} = \Omega \left( \trans{0}{-1} e^{-i \omega_{\rm mw} t}+\trans{-1}{0} e^{i \omega_{\rm mw} t}\right)  \text{ , } 
\end{align}
where $\Omega $ is the Rabi frequency and $\omega_{\rm mw}$ is the frequency of the driving field.
We add $\hat{H}_{\rm mw}$ to the Hamiltonian in Eq.~(\ref{a28}) and move to the frame rotating at $\omega_{\rm mw}$ to arrive at
\begin{align} \label{a29}
\hat H &=\Delta\tilde{s}^{z}+\Omega \tilde{s}^{x} +\sum^{N}_{i=1} \omega_{i}\: \hat{I}^{z}_{i}+\sum^{N}_{i=1} (\tilde{s}^{z}-\frac{1}{2}\one) \mathbf{A}_{i} \cdot\hat{\bf I}_{i} \nonumber\\ 
&+\sum^{N}_{i< j} b_{ij}\left(\hat{I}_i^z \hat{I}_j^z -\frac{1}{4}(\hat{I}_i^+ \hat{I}_j^- + \hat{I}_i^- \hat{I}_j^+)\right)  \text{ , }
\end{align}
where $\Delta = \omega_e +\delta(\theta) -D(\theta) -\omega_{\rm mw}$ is the detuning and we have dropped the contribution from $\ket{+1}$ in the electron spin dynamics justified through the rotating wave approximation.
We, thus, have introduced the two-level electron spin operators $\tilde{s}^{z}=\frac{1}{2}(\trans{0}{0} -\trans{-1}{-1})$ and $\tilde{s}^{x}=\frac{1}{2}(\trans{0}{-1}+\trans{-1}{0})$ in the subspace spanned by states $ \lbrace \ket{0},\ket{-1}\rbrace$.

In the dressed state basis $\ket{\pm} = (\ket{0}\pm\ket{-1})/\sqrt{2}$ representation and neglecting the fast oscillating terms, the Hamiltonian \eqref{a29} reads
\begin{align} \label{a31}
\hat{H}_{\rm eff} &=\Delta \hat{s}^{x}+ \Omega \hat{s}^{z} +\sum^{N}_{i=1} \acute{\omega}_{i} \hat{I}^{z}_{i} +\sum^{N}_{i=1} (\alpha_{i} \hat{s}^{+}\hat{I}^{-}_{i} +\alpha^{\ast}_{i}\hat{s}^{-}\hat{I}^{+}_{i})\nonumber\\
&+\sum^{N}_{i< j} b_{ij}\left( \hat{I}_i^z \hat{I}_j^z -\frac{1}{4}(\hat{I}_i^+ \hat{I}_j^- + \hat{I}_i^- \hat{I}_j^+)\right) \text{ . }
\end{align}
where $\{\hat{s}^z, \hat{s}^x, \hat{s}^\pm\}$ are the Pauli matrices in the dressed basis representation.
Here, the hyperfine coupling $\alpha_{i}=\frac{1}{4} (A^{x}_{i}+i\,A^{y}_{i})$ and the modified nuclear Larmor frequencies $\acute{\omega} _{i}=\omega_{i}-\frac{1}{2}A^{z}_{i}$ have been introduced.
Note that this last secular approximation is justified when $\lbrace\Omega, \omega_{i} \rbrace\gg g_{ei}$, which can be attained by operating the system at high magnetic fields. 
In this work we assume $\Delta = 0$, which is suited for the polarization transfer, and arrive to the effective Hamiltonian Eq.~\eqref{Heff}.

\bibliography{qip2d}

\end{document}